\documentclass[letterpaper,11pt]{article}   %% LaTeX 2e (preferred)
\usepackage{osajnl2} %% do not use with REVTeX4
\usepackage[draft]{hyperref} %% optional

\begin{document}

\title{From vertical-cavities to hybrid metal/photonic-crystal nanocavities:\\ Towards high-efficiency nanolasers}

%% For REVTeX it is possible to automate superscript and e-mail callouts with the superscriptaddress option; see REVTeX4 documentation.

\author{Se-Heon Kim,$^{1,2,*}$ Jingqing Huang,$^{1,2}$, and Axel Scherer$^{1,2}$}

\address{$^1$Department of Electrical Engineering, California Institute of Technology, Pasadena, CA 91125, USA}
\address{$^2$Kavli Nanoscience Institute, California Institute of Technology, Pasadena, CA 91125, USA}

\email{$^*$seheon@caltech.edu} %% email address is required

\begin{abstract}
We provide a numerical study showing that a bottom reflector is indispensable to achieve unidirectional emission from a photonic-crystal (PhC) nanolaser. First, we study a PhC slab nanocavity suspended over a flat mirror formed by a dielectric or metal substrate. We find that the laser's vertical emission can be enhanced by more than a factor of six compared with the device in the absence of the mirror. Then, we study the situation where the PhC nanocavity is in contact with a flat metal surface. The underlying metal substrate may serve as both an electrical current pathway and a heat sink, which would help achieve continuous-wave lasing operation at room-temperature. The design of the laser emitting at 1.3 $\mu{\rm m}$ reveals that relatively high cavity $Q$ of over 1,000 is achievable assuming room-temperature gold as a substrate. Furthermore, linearly-polarized unidirectional vertical emission with the radiation efficiency over 50\% can be achieved. Finally, we discuss how this hybrid design relates to various plasmonic cavities and propose a useful quantitative measure of the degree of the `plasmonic' character in a general metallic nanocavity.   
\end{abstract}

\ocis{140.5960,   %Semiconductor lasers
140.3945,  %Microcavities 
230.5298,  %Photonic crystals
240.6680.}   %Surface plasmons

\maketitle %% null function with osajnl.sty

\section{\label{sect:intro}Introduction}

Spontaneous emission of a dipole emitter can be altered by the presence of a metallic or dielectric reflector. For instance, the spontaneous emission may be completely inhibited inside an appropriately designed optical cavity. Purcell provided the first quantitative analysis on the emission dynamics of an atomic dipole placed inside a cavity characterized by its quality factor ($Q$) and mode volume ($V$).\cite{Purcell46}  It has been now well established that the effect of the spontaneous emission modification will be more pronounced with the higher $Q/V$ ratio.\cite{Gerard_LT99} 

On the other hand, researchers in the field of semiconductor lasers have searched various ways to achieve the so-called `thresholdless' laser in the context of spontaneous emission control by some form of high $Q/V$ cavities.\cite{Yokoyama92} The concept of photonic-crystal (PhC)\cite{Yablonovitch87, Sajeev87} has revolutionized the development of such high $Q/V$ lasers, enabling further miniaturization in device size and reduction in threshold power. The periodic arrangement of dielectrics can result in a forbidden frequency region within which any electromagnetic mode cannot propagate, a property now called the `photonic band gap (PBG).'\cite{Joannopoulos_book} In fact, the presence of such an energy band gap in a one-dimensional (1-D) periodic structure (Distributed Bragg reflector, DBR)\cite{Yeh76} has been known for some time and already applied to the design of the vertical-cavity surface-emitting laser (VCSEL).\cite{Jewell91} The artificial material possessing PBG enables us to confine electromagnetic energy in a volume smaller than the associated wavelength of light. The first nanocavity laser was achieved based on a thin dielectric membrane with periodically arranged air-holes in two-dimensions.\cite{O_Painter_99} Laser gain was provided by multiple layers of quantum wells embedded in the middle of the slab. Though $Q$ of the initial laser cavity was below 500, much progress has been made toward higher $Q$ and smaller $V$ and this has been a major research topic until very recently.\cite{H_Y_Ryu_03, BSSong05, Deotare09} However, the importance of out-coupling efficiency in these wavelength-scale emitters has for some time been neglected. 

\begin{figure}[t]
\centering\includegraphics[width=12cm]{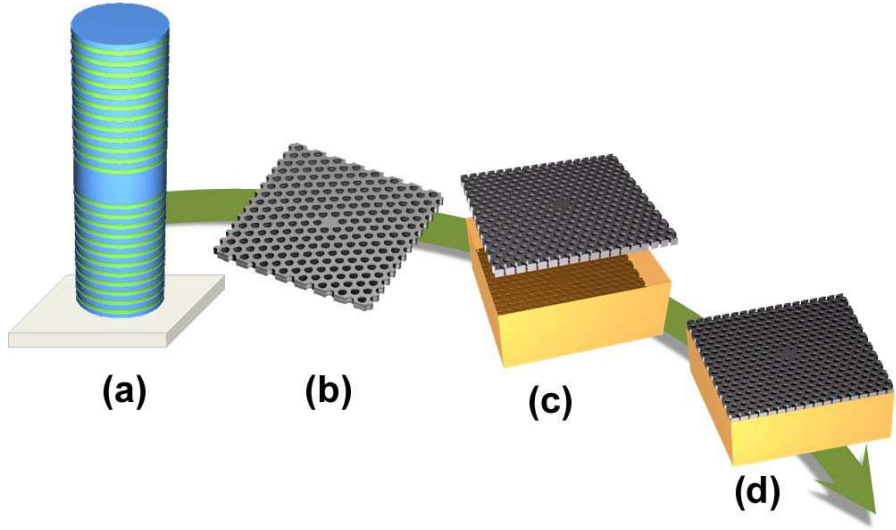} \caption{\label{fig:fig1} Evolution of photonic crystal nanolaser: From vertical-cavity surface-emitting laser to hybrid metal-photonic crystal laser.}
\end{figure}

It is interesting to notice that the basic architecture of PhC nanocavity design has not changed much since the original air-suspended thin-slab geometry.\cite{Painter99} The use of such a thin slab appears to be indispensable to maximize the size of the in-plane PBG.\cite{S_G_Johnson_99} For a high refractive index semiconductor slab, a typical thickness of the PhC slab is 0.5 $a$, where $a$ is the lattice constant. Yet in reality, most PhC slab nanocavities are suspended over a dielectric (or metallic) substrate, as depicted in Fig.~\ref{fig:fig1}(c). Thus, they are no longer isolated but experience feedback from their environment. It has been pointed out that even cavities with $Q$ in the range of 50,000 can `see' the underlying mirror surface, leading to severe modification in the far-field radiation profile.\cite{Kim_PRB_06,Khankhoje10} This phenomenon, as is to be discussed in the following section, has a strong analogy to the aforementioned cavity quantum electrodynamics (QED) example of the point dipole source near a plane mirror.\cite{Hinds} We will show that the combination of a highly reflective bottom mirror and a proper gap size between the PhC slab and the mirror can lead to enhanced far-field directionality from the PhC cavity by more than a factor of 6 in comparison with the PhC nanocavity in the absence of the bottom reflector. This result should be of practical importance for the various PhC based light emitters including nanolasers and single-photon sources.\cite{S_H_Kim_07a, Toishi09, Khankhoje10}

%%%%%%%%%%%%%%%%%%%%%%%%%%%%%%%%%%%%%%%%%%%%%%%%%%%%%%%%%%%%
%%%%%%%%%%%%%%%%%%%%%%%%%%%%%%%%%%%%%%%%%%%%%%%%%%%%%%%%%%%%%

Recently, the realization of an {\em electrically-pumped} nanolaser has drawn renewed attention of many researchers. The first current injection PhC nanolaser was demonstrated in 2004, in which a submicron-sized dielectric post was introduced right below the laser cavity as a means to deliver and confine electrical current.\cite{H_G_Park_04, MKSeo07} However, in order to obtain reasonably high $Q$ for lasing, the post structure needs to be made very thin (diameter $<$ 500 nm) and long (typically $1 \mu{\rm m}$). This could lead to unusually high electrical resistance of over $1 {\rm k}\Omega $, which has been a major bottleneck to achieve continuous-wave (CW) operation of a laser at room-temperature (RT). To circumvent this issues, laterally-doped {\em p-i-n} structures were proposed recently.\cite{Okumura09, Ellis11, Choquette10} CW lasing has been achieved by external cooling at an ambient temperature of $\sim$ 150 K.\cite{Ellis11} Note that {\em optically-pumped} RT-CW lasing operation of a PhC nanolaser was already demonstrated by Nozaki, {\em et al.},\cite{Nozaki07} however, realization of its electrical counterpart has been a severe challenge. 

Therefore, we believe now is the time to reconsider the PhC slab design itself. The concept of using a bottom mirror can naturally lead to the following question -- What would happen if we make the air-gap size zero by placing the whole PhC slab directly on the mirror? We show that a flat metal substrate (See Fig.~\ref{fig:fig1} (d)) can be used effectively to achieve good vertical confinement of the cavity mode.\cite{Kim_ICTON_09} Furthermore, metals are very good conductors for both electrical current and heat dissipation. This may imply that the aforementioned difficulties in building current-injection PhC nanolasers could be mitigated using this hybrid metal-PhC slab design. We shall provide numerical simulation results on the metal-bonded PhC nanocavities in Sec.~\ref{sec:on_metal}. Since the PhC structure is now in contact with the metal substrate, we can expect certain `plasmonic' effects in this hybrid design. In fact, it may not be easy to distinguish between `photonic' and `plasmonic' when both characters coexist in the same structure. Therefore, we will discuss the relationship of this newly proposed design to various `plasmonic' cavities. We will propose a quantitative measure of the degree of the `plasmonic' character, `{\em plasmonicity}', which would provide a useful guideline in the design of `plasmonic' cavities. 

\section{\label{sec:sec1} A PhC nanocavity nearby a bottom reflector}

\subsection{\label{sec:subsec11} Cavity QED analogy}

%%%%%%%%%%%%%%%%%%%%%%%%%%%%%%%%%%%%%%%%%%%%%%%%%%%%%%%%%
\begin{figure}
\centering\includegraphics[width=13cm]{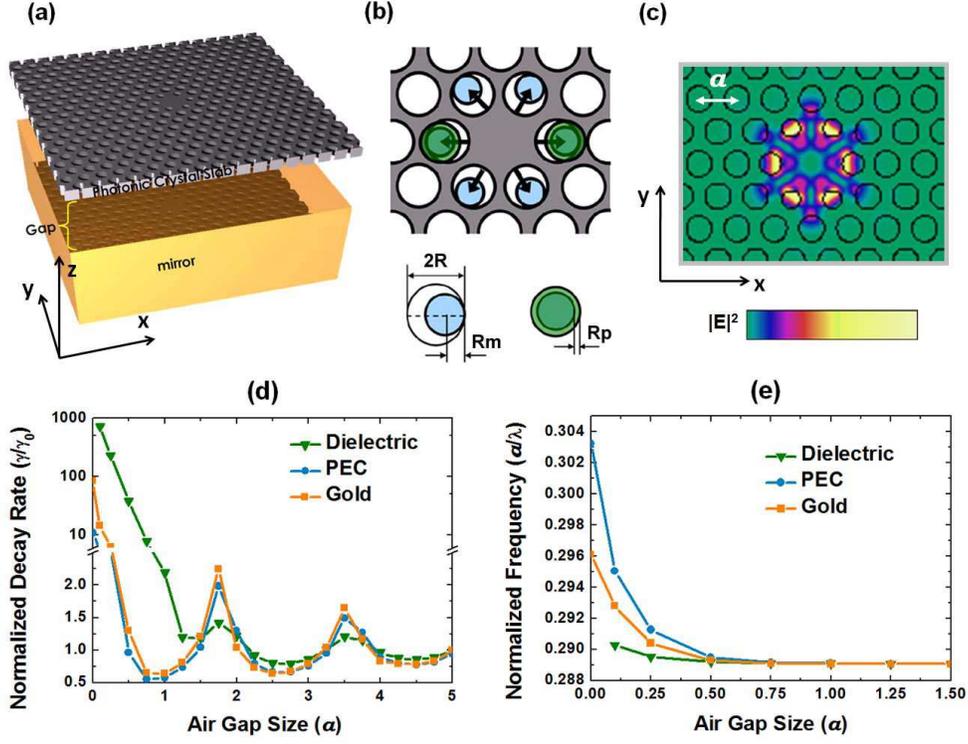}
\caption{\label{fig:fig2}(a) A photonic crystal nanocavity is suspended
   in air above a flat mirror (a bottom reflector). The radiative decay rate ($\gamma$)
   of the nanocavity mode can be tuned as a function of the air-gap size. (b) The design of the photonic crystal nanocavity. Here, two air-holes facing each other are enlarged by $Rp = 0.05 a$. Other parameters are as follows: the slab thickness ($T$) = $0.9 a$, the modified hole radius ($Rm$) = $0.25 a$, and the background hole radius ($R$) = $0.25 a$.  The lattice constant of the photonic crystal is denoted as '$a$' throughout this paper. (c) Electric-field intensity distribution ($|{\bf E}|^2$)
   of the deformed hexapole mode detected in the middle of the slab ($z$ = 0).
   (d) Normalized decay rates ($\gamma / \gamma_0$) of the deformed hexapole mode as a function of the air-gap size. Perfect electric conductor (PEC), gold, and a dielectric of the same refractive index of 3.4 as the slab material are considered as a bottom reflector. In the case of a gold mirror, we assume emission wavelength to be $\sim 1.3 \mu{\rm m}$ with $a$ = 450 nm. Drude model parameters are as follows; $\epsilon_{\infty} = 10.48$, $\omega_p = 1.38 \times 10^{16} {\rm rad/s}$, and $\gamma_m = 1.18 \times 10^{14} {\rm rad/s}$. (e) The resonance frequency also changes as we vary the air-gap size.}
\end{figure}
%%%%%%%%%%%%%%%%%%%%%%%%%%%%%%%%%%%%%%%%%%%%%%%%%%%%%%%%%%

Before going into the details of the hybrid metal-PhC nanocavity, let us first consider a PhC nanocavity suspended over a flat mirror as depicted in Fig.~\ref{fig:fig2}(a). The situation is directly analogous to a point dipole emitter near a plane mirror, which is a well-known cavity QED example.\cite{Hinds} Certainly, there are differences between the PhC nanocavity and the dipole emitter. The PhC nanocavity is much bigger in size than the atomic dipole emitter. The radiation pattern from this hypothetical emitter does not resemble to that of the simple dipole emitter. In general, the complete description requires higher-order multipoles including both electric- and magnetic-multipoles.\cite{M_L_Povinelli_03} Thus, we can view this generalized light emitter as a sum of point-like multipole emitters. Similar modifications in the decay rate and the radiation pattern from the PhC nanocavity are expected. In the absence of a bottom mirror, we find that $Q_0$ of the deformed hexapole mode (Fig.~\ref{fig:fig2}(b)) is $\sim$ 15,000 (subscript $0$ for the cavity with no bottom reflector).\cite{Kim_PRB_06} To make our analysis more analogous to\cite{Hinds}, we translate $Q_0$ into the radiative decay rate using the relation $\gamma_0 = \omega/Q_0$. 

We have performed 3-D finite-difference time-domain (FDTD) simulations to study how $\gamma$ changes in the presence of a bottom reflector. In Fig.~\ref{fig:fig2}(d), we plot normalized decay rate ($\gamma/\gamma_0$) of the deformed hexapole mode as a function of the air-gap size ($d$). Three different types of the bottom reflector are considered in this study -- perfect electric conductor (PEC), gold, and dielectrics. 

First, let us focus on the ideal mirror case (PEC). As expected, we can observe modulations of $\gamma$ as a function of $d$. Both enhanced decay ($\gamma/\gamma_0$ $>$ 1) and suppressed decay ($\gamma/\gamma_0$ $<$ 1) can occur depending on a specific $d$ value. Interestingly, we have found that the modulating features appear to have a certain periodicity. For example, the two consecutive peak positions (one at 1.75 $a$ and the other at 3.5 $a$) are separated by $\sim 0.5 \lambda$. Here, it should be noted that, $1 a$ is equal to $\sim 0.29 \lambda$ for the cavity displaced sufficiently far from the underlying mirror, where $\lambda$ is the emission wavelength measured in vacuum. The observed $\sim 0.5 \lambda$ periodicity reminds us of the two-beam interference condition, where the bottom reflector can reverse the downward propagating beam to the upward direction to produce the interference. We will develop a more rigorous model in the following subsection~\ref{sec:subsec12}. 

Second, the case of a gold reflector is considered. We use the dielectric function of gold at RT for emission wavelength of $\sim$ 1.3 $\mu {\rm m}$.\cite{Johnson72} Slightly increased decay rates in comparison with the previous ideal mirror case are partly due to the additional absorption and the slightly lower reflectivity of gold. The absorption effect becomes more severe for smaller $d$. Indeed, the decay rate becomes noticeably different from the PEC case when $d < 0.5 a$. 

Third, when we replace the gold mirror with a simple dielectric substrate with refractive index of 3.4, the periodic modulation of $\gamma$ becomes much weaker than the previous two cases. This is not surprising because the mirror reflectivity is now only about 30\%. For $d < 1.2 a$, we observe much more enhanced decay rate --- at $d = 0.5 a$, $\gamma/\gamma_0$ is about 38. Such a dramatically enhanced decay rate is mostly due to the enhanced tunneling loss through the bottom substrate and the TE-TM coupling loss in the horizontal direction.\cite{Kim_PRB_06} In fact, light confinement mechanism in the in-plane directions is no longer perfect as we break the vertical symmetry of the PhC slab.\cite{S_G_Johnson_99} 

In all three cases, we observe the break-down of the PBG as $d \rightarrow 0$. However, it is much more severe in the case of a dielectric mirror. We would like to note that $\gamma/\gamma_0$ converges to 1 in the opposite limit of $d \rightarrow \infty$, though it has not been shown explicitly in those plots. Finally, in Fig.~\ref{fig:fig2}(e), it can be seen that the cavity resonance blue-shifts ($\triangle \omega_{\rm blue}$) as $d$ decreases. Such energy-level shifts are also observed from the aforementioned cavity QED example.\cite{Hinds} One may develop a similar perturbative approach based on electric- and magnetic-multipoles to explain the observed $\triangle \omega_{\rm blue}$. However, arguments based on the electromagnetic variational theorem\cite{Joannopoulos_book} would suffice to explain differences in $\triangle \omega_{\rm blue}$ in the three different cases --- the dielectric mirror case shows the smallest $\triangle \omega_{\rm blue}$ due to the more efficient overlap of the electric-field energy with the dielectric mirror region. 

\subsection{\label{sec:subsec12} Enhancing energy directionality: Planewave interference model}

As mentioned previously, the interference of electromagnetic waves is mediated by the bottom mirror to produce the observed decay rate modulation. Remember that the optical loss of the nanocavity is closely related to its far-field radiation pattern, therefore, we expect that the far-field radiation pattern should undergo similar modifications. By using 3-D FDTD, we can directly simulate a far-field radiation pattern, $dP(\theta, \phi)/d\Omega$, of the PhC nanocavity, which represents emitted power ($dP$) within a unit solid angle ($d\Omega$). There exists corresponding wavevector component ${\bf k}$ for each angular direction $(\theta, \phi)$. Thus, any radiation pattern can be decomposed in terms of planewaves $\sum_{\bf k} (dP/d\Omega) ({\bf k})$. This planewave decomposition provides an alternative way to the multipole expansion method. From now on, we will focus on planewaves with $k_x = 0$ and  $k_y = 0$, since we are interested in the vertical directionality of laser emission.

\begin{figure}%[b]
\centering\includegraphics[width=12cm]{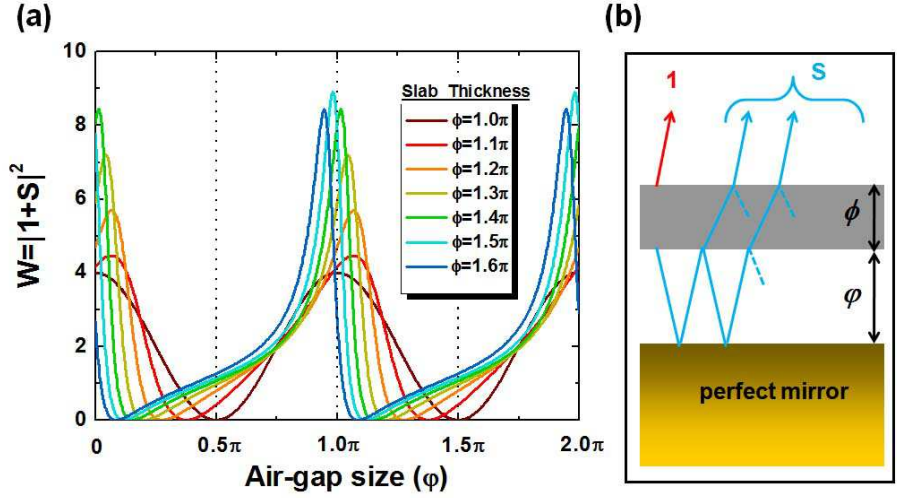} \caption{\label{fig:fig3}(a) Vertical emission
enhancement factor ($W$) obtained by the planewave interference model. Both the air-gap size and the slab thickness are varied and the results of varying slab thicknesses were shown as multiple curves as a function of the air-gap. We have assumed the effective refractive index of the slab ($n_{eff}$) to be 2.6, which will result in $r_0=(n_{eff}-1)/(n_{eff}+1)\approx 0.44$. (b) A schematic of the model for the photonic crystal nanocavity suspending over a bottom reflector. The perforated slab is replaced with a uniform dielectric slab with $n_{eff}$ and the underlying mirror is assumed to be PEC.}
\end{figure}

Figure~\ref{fig:fig3}(b) describes how the complex 2-D slab nanocavity is simplified in the spirit of the planewave decomposition. The perforated PhC membrane is approximated as a uniform dielectric slab with an effective refractive index.\cite{Fan02} We will deal with the PEC mirror case only, where any wave incident upon it will undergo a $\pi$-phase shift. Detailed calculations have been described in our earlier publication,\cite{Kim_PRB_06} here we would like to summarize the essential ideas. First, the final result will depend only on the effective index of the slab, $n_{\rm eff}$, the phase thickness of the slab, $\phi$, and the phase thickness of the air-gap, $\varphi$. Once $n_{\rm eff}$ is determined, we can derive coefficients of amplitude reflection and transmission for a single dielectric interface such that $r_0 = (n_{\rm eff}-1)/(n_{\rm eff}+1)$ and $t_0 = (1- r_0^2)^{1/2}$. Second, all of the scalar waves will be treated as complex numbers in the form $\exp(ik_z z -i\omega t)$. Any wave traveling across a phase thickness of $\phi_0$ will gain the same amount of phase. Third, we assume that there are two `seed' waves, one on the top of the slab and one at the bottom of the slab. Both waves are assumed to propagate in the opposite directions with the equal amplitude and phase. This assumption can be justified by the fact that the original PhC nanocavity mode is symmetric with respect to the $z$ = 0 plane. Fourth, the wave initially propagating in the downward direction will be redirected upward by the bottom mirror. In Fig.~\ref{fig:fig3}(b) $S$ denotes the sum of all such waves finally detected in the far-zone at $\theta = 0$. During this complex process, a fraction of the energy can couple to the PhC nanocavity mode. If this coupling occurs, then the resonant mode can produce wavevector components with $k_x \neq 0$ or  $k_y \neq 0$. In this model, we assume that this coupling process is negligible and the in-plane wavevector components are conserved. This assumption is analogous to the {\em weak-coupling regime} in the cavity QED\cite{Hinds} --- once a photon leaves the emitter (the PhC resonant mode), it does not strongly interact with the original emitter. The photon will `see' only mirror boundaries of an external cavity (the bottom mirror and the uniform dielectric slab) whose $Q$ is much smaller than that of the emitter. It should be noted that similar approaches have been applied to the calculation of spontaneous emission inhibition and enhancement of a dipole emitter in a cavity.\cite{Dowling91}

%%%%%%%%%%%%%%%%%%%%%%%%%%%%%%%%%%%%%%%%%%%%%%%%%%%%%%%
%%%%%%%%%%%%%%%%%%%%%%%%%%%%%%%%%%%%%%%%%%%%%%%%%%%%%%%%

With all the assumptions above, the relative vertical enhancement can be written as $|1+S|^2$, which is normalized to the result in the absence of the bottom reflector. Here is the final result of the summation. 

\begin{equation}
W \equiv \left| 1 + S \right|^2 =  \left| 1 + \frac{t_0^2 e^{i \phi}}{ (1-r_0^2 e^{2 i \phi})(r_0 -
e^{-2i\varphi})-r_0 t_0^2 e^{2 i \phi} } \right|^2.
\end{equation}

We plot $W$ as a function of the air-gap size ($\varphi$), where we assume $n_{\rm eff}$ to be 2.6 in consideration of the effective surface coverage ratio of $\sim 1 - [2 \pi R^2 / (\sqrt{3} a^2)]$ for the PhC slab with $R = 0.35 a$. Fig.~\ref{fig:fig3} shows the results for various $\phi$ values (the slab thickness). When $\phi$ satisfies the `slab resonance' condition ($\phi = m \times \pi$, where $m$ is an integer), the slab will be `transparent' 
for any wave incident upon it. Under this condition, the redirected wave from the bottom mirror does not `see' the slab, which result in the simple two-beam interference. $W$ modulates between 0 and 4 by varying the air-gap size. The effective thickness of a PhC slab is typically chosen to be around `$T \approx \lambda/(2 n_{\rm eff})$' in order to maximize the PBG,\cite{S_G_Johnson_99} which happens to be near the slab resonance condition. This implies that near 100\% transmission is possible by slightly tuning the emission wavelength toward the exact slab resonance condition.\cite{Fan02} Then, the subsequent optimization of the air-gap size can bring up the vertical emission intensity by up to a factor of 4. Here, it is worth mentioning that the enhancement factor is a quantity proportional to the decay rate. Thus, the fact that $ W_{\rm max} = 4 > 1+1$ does not violate the conservation of energy --- most of the electromagnetic energy, $U(t)$, is stored in the cavity and it will experience exponential decay in time such that $U(t) = U(t_0) \exp [\gamma (t-t_0)]$, where the decay rate, $\gamma$, is proportional to $W$. Conversely, if the air-gap size is chosen to be $1.5\pi$, then, for an ideal 1-D system, we can completely quench the radiation to the vertical direction, resulting in infinitely large $Q$. Similar unbounded $Q$ behavior has been found from the second $\Gamma$-point band-edge state in a 2-D honeycomb-lattice PhC slab\cite{Bakir06} and the $\Gamma$-point band-edge state in a 2-D triangular-lattice PhC slab,\cite{Ochiai01} with and without using a bottom reflector, respectively.

Now let us turn to our original problem and find out the condition that maximizes $W$. Surprisingly, $W$ can increase up to about 9 by tuning the slab thickness near the `slab antiresonance' ($\phi = (m + 0.5) \times \pi$) as shown in Fig.~\ref{fig:fig3}(a).\cite{Kim_PRB_06} This phenomenon can be understood by the fact that the slab becomes highly reflective (reflectance can be as high as 70\% due to the absence of the slab resonance).\cite{Fan02} Thus, the slab and the bottom mirror system constitutes a certain vertical resonator, which enhances the density of photon modes near $k_x = k_y = 0$ point. In the following subsection, through rigorous 3-D FDTD simulations, we will show that over six-fold vertical emission enhancement can be achieved in the deformed hexapole mode example. 

%%%%%%%%%%%%%%%%%%%%%%%%%%%%%%%%%%%%%%%%%%%%%%%%%%%%%
\begin{figure}
\centering\includegraphics[width=12cm]{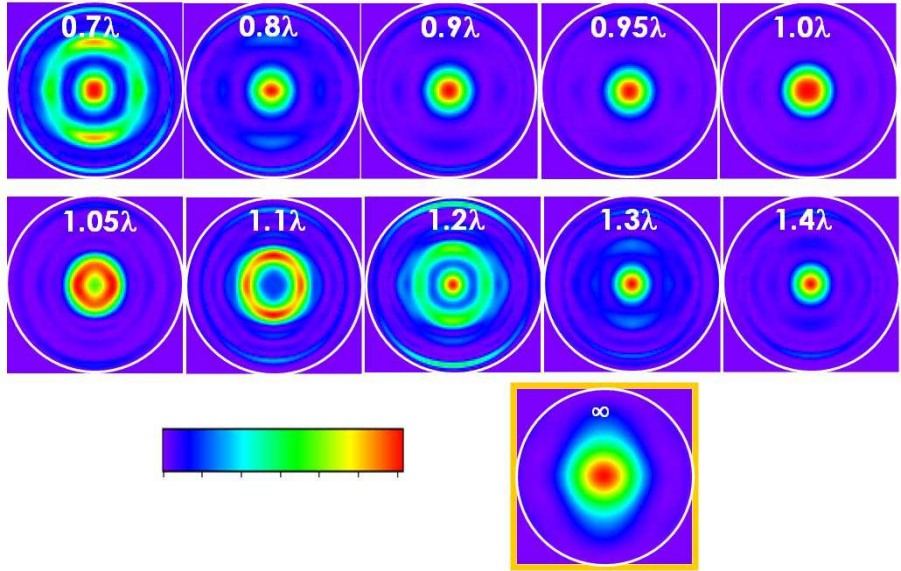} \caption{\label{fig:fig4}FDTD
simulated far-field emission profiles from the deformed hexapole mode shown in Fig.~\ref{fig:fig2}. Far-field patterns detected over the hemispherical surface are transformed into the 2-D plane by using a simple mapping defined by $x=\theta \cos \phi$ and $y= \theta \sin \phi$. Numbers represent the air-gap size normalized to the emission wavelength of the reference cavity ($\infty$) in the absence of the mirror.}
\end{figure}
%%%%%%%%%%%%%%%%%%%%%%%%%%%%%%%%%%%%%%%%%%%%%%%%%%%%%

\subsection{\label{sec:subsec13} Enhancing energy directionality: FDTD}
\begin{figure}
\centering\includegraphics[width=12cm]{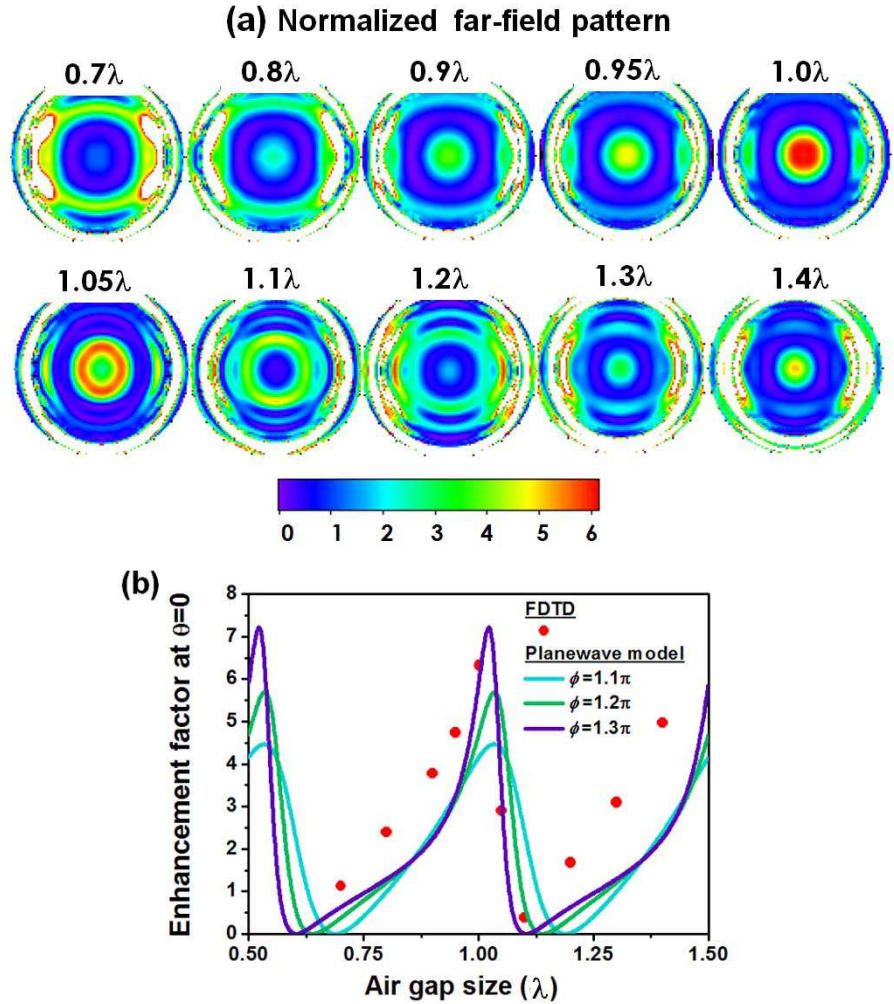}
\caption{\label{fig:fig5}(a) The far-field emission profiles shown in Fig.~\ref{fig:fig4} are normalized by the reference far-field pattern ($\infty$ in Fig.~\ref{fig:fig4}), where white regions denote values $>$ 6.33. (b) We extract $\theta = 0$ components from the normalized far-field patterns and plot them together with the theoretical curves obtained by the planewave interference model. }
\end{figure}

The planewave interference model predicts that vertical emission can be enhanced by more than a factor of 4. Here, we would like to test the validity of the model by using rigorous FDTD simulation. In our previous paper, we have adopted the near- to far-field transformation algorithm for more efficient and fast simulation.\cite{Kim_PRB_06} However, for more accurate results, we will perform direct FDTD simulations in a very large computational domain with $(L_x \times L_y \times L_z) \geq (6 \lambda \times 6 \lambda \times 6 \lambda)$. We take the same deformed hexapole mode shown in Fig~\ref{fig:fig2} as an example and assume a PEC bottom mirror. Note that we will vary only the air-gap size while we keep the slab thickness constant at $0.9 a$. First, far-field emission profiles, $f(\theta, \phi)$, are obtained by detecting the radial component of the Poynting vectors over a hemispherical surface whose radius is larger than $3 \lambda$ (See Fig.~\ref{fig:fig4}). These far-field patterns are then normalized by the original far-field pattern, $f_0(\theta, \phi)$, obtained in the absence of the bottom reflector (See Fig.~\ref{fig:fig4}). By varying the gap size, it is found that the internal electromagnetic energy, $U(t)$, does not show noticeable change within the FDTD time needed for the far-field simulation. Thus, errors associated with small variations in $U(t)$ will be less than 1\%. We present normalized far-field patterns, $\tilde{f}(\theta, \phi) \equiv f/f_0$, in Fig.~\ref{fig:fig5}(a), where we use a simple mapping defined by $x=\theta \cos \phi$ and $y= \theta \sin \phi$. Thus, the center corresponds to $\theta = 0$ and the angle $\theta$ will be proportional to the radial distance from the origin. Clearly, we can have more than six-fold enhancement by choosing the air-gap to be near $1 \lambda$. 

In Fig.~\ref{fig:fig5}(b), we plot $\tilde{f}(\theta=0)$ as a function of air-gap size. The three solid lines are from the previous planewave interference model, for $\phi = 1.1\pi, 1.2\pi$, and $1.3\pi$, respectively. Generally, the curve for $\phi = 1.3 \pi$ shows good agreement with the FDTD results. Interestingly, the model also predicts the asymmetric behavior around $1.0 \lambda$ as is confirmed by the FDTD result --- the enhancement value decreases more slowly on the left side than on the right side. Thus, both the FDTD result and the planewave interference model confirm that the vertical emission can be enhanced by more than a factor of 6 by employing a highly reflective mirror at the bottom.

\subsection{Applications}

These findings will be of significant importance to the design of various PhC based light emitters, such as nanolasers and single-photon sources.\cite{S_H_Kim_07a,Toishi09, Khankhoje10} The emission wavelength is a rather fixed property by the active material embedded in the slab. In most wafer designs, the final selective wet-etching process will leave a flat surface of a dielectric substrate, which will serve as a good bottom mirror as we have shown in Fig.~\ref{fig:fig2}. Furthermore, in GaAs material systems, we may include a GaAs/AlAs DBR whose resultant reflectivity will be over 98\%. To optimize the vertical directionality, the following two criteria should be considered for the wafer design. 1) The thickness of the sacrificial layer (which is to become the air-gap) should be designed to be equal to the emission wavelength. 2) The PhC slab thickness should be chosen to be near the slab antiresonance. However, in consideration of the PBG size, the slab thickness may be limited by $1.0 a$ ($\phi_{\rm max} \approx 1.3 \pi$). It should be noted that, the enhanced directional emission has been verified from InP/InGaAsP PhC nanolasers by direct measurement of far-field emission profiles.\cite{J_Kang_09}  

\section{\label{sec:on_metal} A PhC nanocavity on a metal substrate}
As mentioned previously, the planewave interference model may not be applicable to the case where the air-gap size is below $\lambda /2$. However, we have seen already what would happen in the limit of zero air gap from Fig.~\ref{fig:fig2}(d,e). In the case of the dielectric mirror, we cannot find any mode whose $Q$ is higher than 30. However, a relatively high-$Q$ ($> 1,000$) mode is found by assuming the PEC mirror. The case of the gold mirror is not as good as the case of the PEC mirror --- $Q \sim$ 150 has been obtained. Though this $Q$ value is already comparable with typical $Q$ values from previously reported metallic nanocavities,\cite{Min09,Oulton09,Seo_NL09,Hill_OPEX_09,J_Huang_10} we will propose a method to bring it up to several thousands. In this section, we will explore another opportunity with this new metal-PhC nanocavity design (Fig.~\ref{fig:fig6}(a)) for a practical nanolaser. The fact that a metal substrate is in direct contact with the PhC slab structure may mitigate aforementioned difficulties in building electrically-pumped nanolasers, namely, excess electrical and thermal resistance. Moreover, the metal may work as a good bottom reflector so that the vertical directional emission could be enhanced. This hybrid metal-PhC design will also be an important building block in the field of plasmonics in the context of building efficient plasmonic lasers.\cite{Hill_JOSAB_10}

%%%%%%%%%%%%%%%%%%%%%%%%%%%%%%%%%%%%%%%%%%%%%%%%%%%%%%%%%
%%%%%%%%%%%%%%%%%%%%%%%%%%%%%%%%%%%%%%%%%%%%%%%%%%%%%%%%%%

\subsection{$Q$, $V$, and Purcell factor $F_p$} 
\begin{figure}
\centering\includegraphics[width=14.5cm]{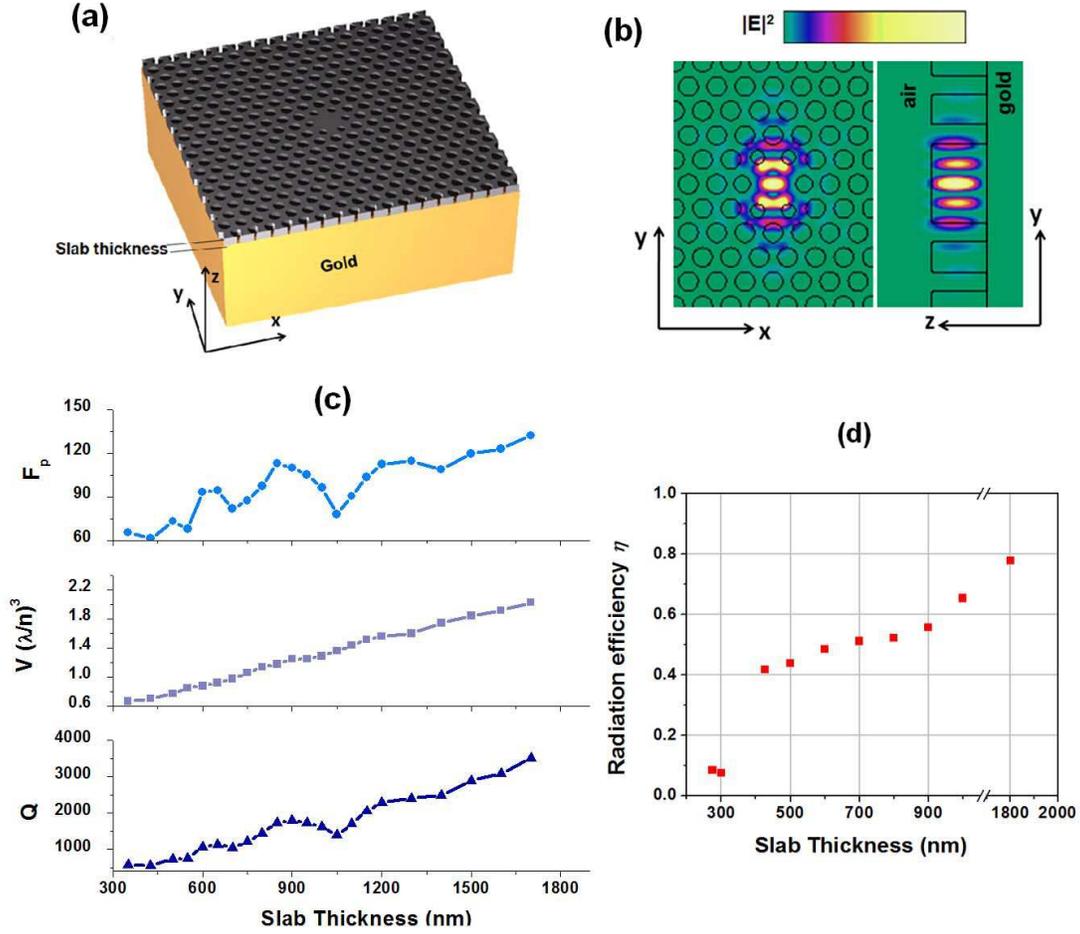}
\caption{\label{fig:fig6}(a) A PhC nanocavity is brought into contact with the underlying metal substrate. We assume realistic optical constants of gold at room-temperature, which is implemented using the single-pole Drude model in FDTD. (b) FDTD simulated electric-field intensity profiles ($|{\bf E}|^2$) of the dipole mode when the slab thickness is 606 nm. Other structural parameters are as follows: $Rm$ = $0.25 a$, $R$ = $0.25 a$, $Rp$ = $0.05 a$, and $a$ = 315 nm. (c) Optical properties of the dipole mode. Quality factor ($Q$), effective mode volume ($V$), and Purcell factor ($F_p$) derived from $Q$ and $V$ are plotted as a function of the slab thickness. Here, slightly different lattice constants ($a$) have been used for different slab thicknesses to keep the emission wavelength a constant at $\sim 1.3 \mu{\rm m}$. (d) The total electromagnetic energy contained in the cavity dissipates into two independent loss channels, one in the form of propagating radiation in air and the other in the form of absorption in metal. Here, radiation efficiency refers to the fraction of total dissipation into the radiation.}
\end{figure}

Before beginning with our discussions on the result in Fig.~\ref{fig:fig6}, let us clarify the definitions of $Q$, $V$, and $F_p$ and several energy related quantities, especially in the context where dispersive metals are involved. 

$Q_{\rm tot}$ is defined through the decay rate of the total electromagnetic energy contained in the cavity, $U_{EM} (t)$, such that 
\begin{equation}
U_{EM} (t) = U_{EM} (0) \exp \left[ - \frac{\omega}{Q_{\rm tot}} t \right]\label{eq:eq3}.
\end{equation} 
Here, $U_{EM} (t)$ is the sum of the electric-field energy, $U_E (t)$, and the magnetic-field energy, $U_M (t)$, stored in the PhC nanocavity mode. $U_E (t)$ and $U_E (t)$ can be defined by the energy density functions, $u_E({\bf r}, t)$ and $u_M({\bf r}, t)$, respectively. 
\begin{eqnarray}
U_E (t) &\equiv& \int_V d^3 {\bf r} ~ u_E({\bf r}, t)\label{eq:eq4a} \\
U_M (t) &\equiv& \int_V d^3 {\bf r} ~ u_M({\bf r}, t)\label{eq:eq5a}
\end{eqnarray}
In the case where dispersive media are involved in the energy calculations, special care must be taken. In this paper, we will use the following definitions for $u_E$ and $u_M$, as has been noted by Chang and Chuang.\cite{Chang09_QE, Chang09_OL}
\begin{eqnarray}
u_E({\bf r}, t) &\equiv& \frac{\epsilon_0}{2} {\rm Re} \left[ \frac{d(\omega \epsilon )}{d \omega} \right]  \left< {\bf E}({\bf r}, t) \cdot {\bf E}({\bf r}, t) \right>_T\label{eq:eq4} \\
u_M({\bf r}, t) &\equiv& \frac{\epsilon_0}{2} {\rm Re} \left[ \epsilon(\omega) \right] \left< {\bf E}({\bf r}, t) \cdot {\bf E}({\bf r}, t) \right>_T\label{eq:eq5}
\end{eqnarray}
Here, the bracket $\left< \cdot \cdot \cdot \right>_T$ denotes time average over one optical period. If ${\bf r}$ lies in a normal dielectric medium of $\epsilon_d$, then the above expressions will take the familiar forms of $u_E ({\bf r}, t) = u_M ({\bf r}, t) = (\epsilon_0 \epsilon_d /2) \left< {\bf E}({\bf r}, t) \cdot {\bf E}({\bf r}, t) \right>_T$, where the equality between the electric-field energy and the magnetic-field energy comes from the fact that the resonant mode is harmonically oscillating in time.\cite{Jackson_book}
 
If ${\bf r}$ lies in a Drude medium where $\epsilon_m (\omega) = \epsilon_{\infty} - \omega_p^2/(\omega^2 + i \gamma_m \omega)$, we obtain 

\begin{eqnarray}
u_E({\bf r}, t) &=& \frac{\epsilon_0}{2} \left\{ \epsilon_{\infty} + \frac{\omega_p^2 (\omega^2 - \gamma_m^2)}{ (\omega^2 + \gamma_m^2)^2 }   \right\} \left< {\bf E}({\bf r}, t) \cdot {\bf E}({\bf r}, t) \right>_T\label{eq:eq6} \\
u_M({\bf r}, t) &=& \frac{\epsilon_0}{2} \left\{ \epsilon_{\infty} - \frac{\omega_p^2}{ \omega^2 + \gamma_m^2}   \right\} \left< {\bf E}({\bf r}, t) \cdot {\bf E}({\bf r}, t) \right>_T\label{eq:eq7}
\end{eqnarray}

Throughout this paper, we adopt the cavity QED definition for the mode volume $V$, which, using the above definition of $U_{EM}$, can be written as 
\begin{equation}
V \equiv \frac{ U_{EM} (t) }{ {\rm max} \{ U_{EM} (t) \} } \label{eq:eq11}
\end{equation} 

Using $Q_{\rm tot}$ (Eq.~\ref{eq:eq3}) and $V$ (Eq.~\ref{eq:eq11}), Purcell's figure of merit can be written as\cite{Gerard_LT99}
\begin{equation}
F_p \equiv \frac{3 Q_{\rm tot}}{4 \pi^2 V} \left( \frac{\lambda}{n} \right)^3,\label{eq:eq12}
\end{equation} 
where a single emitter is assumed to be located in a medium of a refractive index $n$.

%%%%%%%%%%%%%%%%%%%%%%%%%%%%%%%%%%%%%%%%%%%%%%%%%%%%%%%
%%%%%%%%%%%%%%%%%%%%%%%%%%%%%%%%%%%%%%%%%%%%%%%%%%%%%%%%%

Now consider the dipole mode in the PhC cavity in contact with the gold substrate as shown in Fig.~\ref{fig:fig6}(a,b). As mentioned previously, $Q_{\rm tot}$ cannot be larger than 200 in the case of the deformed hexapole mode in a thin slab (slab thickness $T = 0.9 a$). Traditionally, the PhC slab thickness is not chosen to be much larger than $1.0 a$, because the in-plane PBG begins to close at this value. However, Tandaechanurat, {\em et al.} have shown recently that relatively high $Q$ can be obtained from the PhC dipole mode suspended in air after the PBG closure.\cite{Tandaechanurat08} They have found that $Q_{\rm tot}$ can increase up to $\sim 10,000$ at $T \sim 1.4 a$. 
In Fig.~\ref{fig:fig6}(c), we plot $Q_{\rm tot}$ and $V$ as a function of the slab thickness. We have increased the slab thickness up to $6 a \sim 1,800$ nm. Surprisingly, $Q_{\rm tot}$ shows a rather monotonic increase and a value of 3,000 can be obtained when $T > 1,500$ nm. $Q_{\rm tot}$ may be further improved by inserting a thin low refractive index layer between the PhC slab and the bottom mirror,\cite{Mizrahi_08} which will reduce the optical overlap with the underlying gold. What is interesting here is that $Q_{\rm tot}$ seems to increase indefinitely with increasing $T$, although the in-plane PBG is completely closed. This unusual behavior may be understood based on the waveguide dispersion along the $z$ direction ($\omega$-$k_z$ diagram) --- the structure can be viewed as a PhC fiber with a finite length.\cite{Joannopoulos_book, Ibanescu05} In the case of a thin slab, $k_z$ is not a well-defined quantum number. However, as $T$ increases, $k_z$ of the fundamental mode can be more and more precisely defined in accordance to the uncertainty relation between $\triangle z$ and $\triangle k_z$. In fact, $k_z$ of the fundamental slab mode will converge to zero in the limit of infinite $T$. This $k_z = 0$ point corresponds to the ideal 2-D limit in the $\omega$-$k_z$ diagram. Thus, the mode can be more and more confined as $T$ increases. It should be noted that, with a certain large $T$, radiation loss will occur mostly in the horizontal directions ($x$-$y$ plane) rather than in the vertical direction ($z$).\cite{SHKim_unplished} This is due to the presence of the zero group velocity dispersion at $k_z =0$ point. The slow group velocity mode will effectively reduce the scattering loss at the top surface of the PhC slab.\cite{Ibanescu05} It turns out that this vertical confinement mechanism works much more effectively than the horizontal confinement mechanism by the PhC mirror after $T > 1 a$, hence more radiation in the horizontal direction. For practical applications, however, we may redirect the horizontal propagating wave's energy into the vertical direction by employing grating couplers.\cite{Taillaert02}

On the other hand, $V$ tends to increase almost linearly with the slab thickness. It should be noted that $V \sim 1.0 (\lambda/n)^3$ at $T \sim 720$ nm, which is comparable with that of the widely used L3 nanocavity (about 1.2 $(\lambda/n)^3$).\cite{BSSong05} We have also estimated the maximum achievable spontaneous emission rate enhancement through the Purcell factor $F_p$.\cite{Gerard_LT99} It is shown that, theoretically, $F_p$ of over 100 can be achieved, a reasonably high value within the weak-coupling regime of the cavity QED.  

\subsection{Threshold gain}
$Q_{\rm tot}$ tells us how much threshold gain will be required to achieve lasing. According to the recent formulation by Chang and Chuang,\cite{Chang09_QE, Chang09_OL} threshold gain, $g_{th}$, is given by 

\begin{equation}
g_{th} = \frac{1}{\Gamma_E \cdot Q_{\rm tot}} \cdot \frac{2 \pi n_{g,a}}{\lambda},\label{eq:eq14}
\end{equation} 
where $n_{g,a}$ is the material group index defined as $\partial[\omega n_a (\omega)]/\partial \omega$ and $n_a(\omega)$ is the refractive index of the active region. $\Gamma_E$ is the energy confinement factor defined as 

\begin{equation}
\Gamma_E = \frac{\int_{V_a} d^3 {\bf r} ~ u_E ({\bf r}, t) + u_M ({\bf r}, t) }{\int_{V} d^3 {\bf r} ~ u_E ({\bf r}, t) + u_M ({\bf r}, t) }. \label{eq:eq15}
\end{equation}
Here, the above volume integration for $V_a$ is taken over the volume of the active region and $u_E$ and $u_M$ are defined in Eqs.~(\ref{eq:eq4}) and~(\ref{eq:eq5}). Strictly speaking, in situations involving dispersive media, $\Gamma_E$ is a quantity that depends on time. However, in most cases, we can safely assume $\Gamma_E$ as a constant for a given resonant mode. 

We take the PhC dipole mode as an example, whose electric-field intensity profiles are shown in Fig.~\ref{fig:fig6}(b). We choose the slab thickness to be 606 nm, roughly $2 a$. We find that $Q_{\rm tot}$ and resonant wavelength are 1,032 and 1307 nm, respectively. Assuming the appropriate number of quantum wells as the gain material, $\Gamma_E$ $\sim$ 10\% is not so difficult to achieve, from which  $g_{th}$ is estimated to be $\sim 1000 {\rm cm}^{-1}$. This $g_{th}$ value is achievable by employing conventional InP/InGaAsP quantum wells.\cite{Fujita99} 

\subsection{Radiation efficiency}
Because of the presence of the absorbing metal layer near the PhC cavity, energy contained in the cavity will be lost by two independent mechanisms: absorption loss in the metal ($\sim 1/Q_{\rm abs}$) and radiation loss into air ($\sim 1/Q_{\rm rad}$). Therefore, the total $Q$ can be decomposed into the radiation $Q$ and the absorption $Q$ in the following manner. 

\begin{equation}
\frac{1}{Q_{\rm tot}} = \frac{1}{Q_{\rm rad}} + \frac{1}{Q_{\rm abs}} \label{eq:eq2}
\end{equation}

We define the radiation efficiency, $\eta_{\rm rad}$, as the ratio of total radiated power over the total dissipated power. This can be written in terms of quality factors in the following way.  
\begin{equation}
\eta_{\rm rad} \equiv \frac{1/Q_{\rm rad}}{1/Q_{\rm tot}} = 1- \frac{1/Q_{\rm abs}}{1/Q_{\rm tot}}\label{eq:eq10}
\end{equation} 
Here, $Q_{\rm tot}$ can be easily estimated through field decay in the time-domain using Eq.~(\ref{eq:eq3}). Estimating $Q_{\rm abs}$ or $Q_{\rm rad}$ requires additional volume integration or surface integration. For example, to calculate absorbed power in the Drude metal, we should use the following volume integration.\cite{Jackson_book} 
\begin{equation}
P_{\rm abs} (t) \equiv \int_V d^3 {\bf r} ~ \omega \epsilon_0 {\rm Im} \left[ \epsilon_m (\omega) \right] \left< {\bf E}({\bf r}, t) \cdot {\bf E}({\bf r}, t) \right>_T\label{eq:eq8}
\end{equation} 
Then, $Q_{\rm abs}$ can be calculated by 
\begin{equation}
Q_{\rm abs} = \omega \frac{ U_{EM} (t) }{ P_{\rm abs} (t) }\label{eq:eq9}
\end{equation} 
On the other hand, $Q_{\rm rad}$ can be calculated by 
\begin{equation}
Q_{\rm rad} = \omega \frac{ U_{EM} (t) }{ P_{\rm rad} (t) },\label{eq:eq13}
\end{equation} 
where $P_{\rm rad} (t)$ is given by 
\begin{equation}
P_{\rm rad} (t) \equiv \oint_S d^2 {\bf r} \cdot \left< {\bf E}({\bf r}, t) \times {\bf H}({\bf r}, t) \right>_T\label{eq:eq14}
\end{equation} 
However, this type of surface integration usually requires a much larger computational domain since the surface of the integration should be located sufficiently far from the near-zone of the PhC nanocavity mode. Thus, we have adopted Eq.~(\ref{eq:eq9}) for the calculation of the radiation efficiency.

As expected, $\eta_{\rm rad}$ is quite low (below 10 \%) when $T \leq$ 300 nm. However, we can bring up $\eta_{\rm rad}$ by simply increasing the slab thickness. It is found that $\eta_{\rm rad}$ becomes about 50 \% at $T = 600$ nm. From this slab thickness, we can argue that the device's vertical radiation efficiency will begin to compete with the traditional PhC slab cavity suspended in air, by which at most 50 \% of photons can be collected from the top (without the bottom mirror). In any case, such $\eta_{\rm rad}$ values will be much higher than that of other classes of metallic nanocavities, where $Q_{\rm tot}$ tends to be limited by $Q_{\rm abs}$.\cite{Min09,Seo_NL09} 

%%% equation number goes up to 15

\subsection{Far-field emission}

%%%%%%%%%%%%%%%%%%%%%%%%%%%%%%%%%%%%%%%%%%%%%%%%%%%%%%%
\begin{figure}%[htb]
\centering\includegraphics[width=13cm]{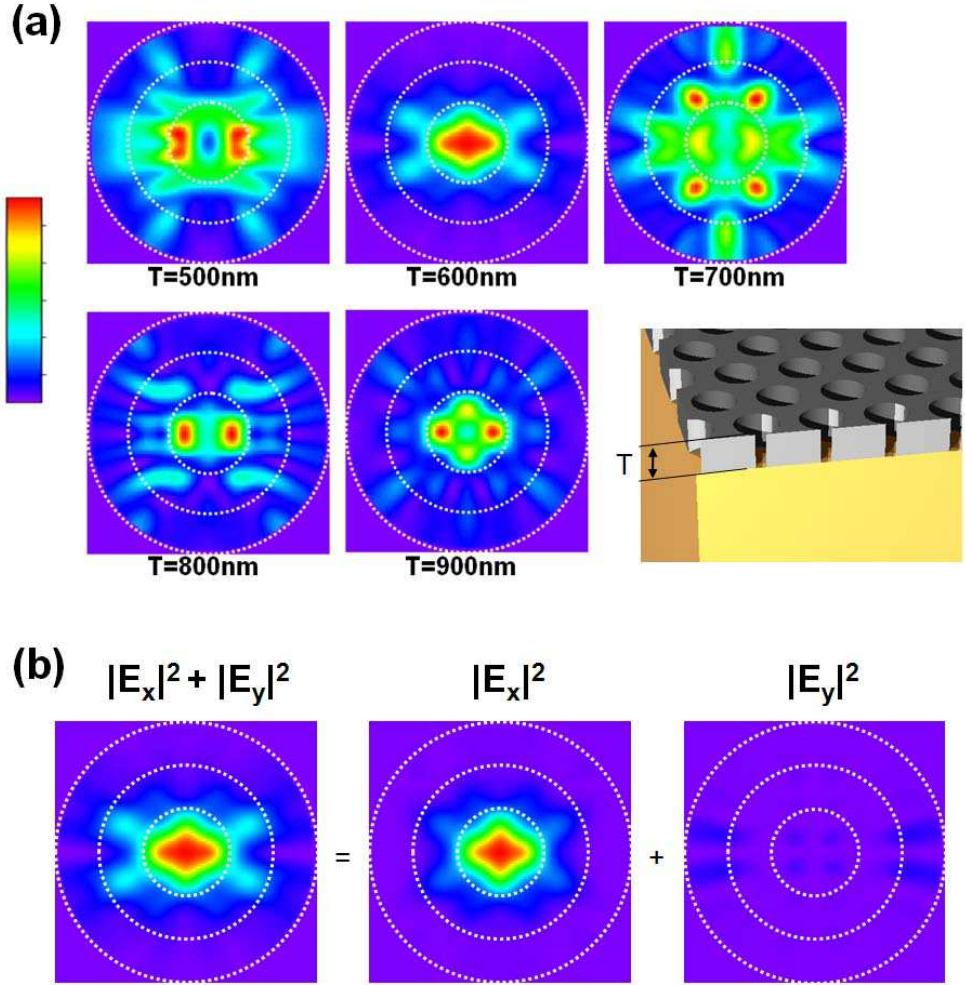}
\caption{\label{fig:fig7}Here, we assume RT gold substrate and analyze the same dipole mode shown in Fig.~\ref{fig:fig6}(b). (a) Far-field emission profiles by varying the slab thickness from 500 nm to 900 nm. (b) Polarization resolved far-field pattern for T = 600 nm. In the simulation, microscopic linear polarizers, one polarized along $x$ direction and the other $y$ direction, are assumed to scan over the hemispherical surface to measure $|E_x|^2$ and  $|E_y|^2$, respectively.}
\end{figure}
%%%%%%%%%%%%%%%%%%%%%%%%%%%%%%%%%%%%%%%%%%%%%%%%%%%%%%%

Far-field directionality can be tuned by varying the slab thickness. Similar systematic optimization of the vertical collection efficiency has been reported for the nanowire cavity sitting on a flat metal surface.\cite{Friedler09} For such a simple cavity geometry, a Fabry-P\'{e}rot model can be used to optimize the far-field directionality. However, our PhC nanocavity involves complex geometrical features together with the zero group velocity dispersion along the $z$ direction, making it difficult for us to develop a simple model as has been done for the PhC nanocavity suspended over the bottom mirror. Therefore, we have used 3-D FDTD and the near- to far-field transformation algorithm developed in our previous work.\cite{Kim_PRB_06} 
Fig.~\ref{fig:fig7} shows the evolution of far-field patterns as a function of the slab thickness. The result of $T$= 600 nm looks most promising; about 50 \% of photon emitted to the top surface will be collected within $\pm 30^{\circ}$. 
Fig.~\ref{fig:fig7}(b) shows that the vertical emission is linearly polarized along the $x$ direction, whose direction of polarization has been determined by the two enlarged air-hole positions (See Figs.~\ref{fig:fig2}(b) and ~\ref{fig:fig6}(b)). 

\section{\label{sec:plasmonicity} Degree of plasmonic effects: {\em Plasmonicity}}

As discussed in the previous section, the effect of the thick slab on the vertical confinement mechanism can be understood from the viewpoint of a PhC fiber. On the other hand, the effect of the bottom mirror may be understood by the method of image charges.\cite{Jackson_book} In the case of a PEC mirror, this scheme works perfectly --- the fundamental resonant mode (even) in a PhC slab with thickness $T$ is equivalent to the first-order mode (odd) in a PhC slab with thickness $2T$.\cite{Joannopoulos_book} However, in the case of a realistic metal mirror, the situation is not so simple. The existence of the evanescent field of the cavity mode in the metal does not preclude the possibility of interesting {\em plasmonic} effects within such a structure.    

So far, interesting aspects particular to SPP modes have been emphasized by many researchers. Two representative examples are the extremely large local electric-fields in a metal nano-antenna (hot spots)\cite{Li11} and various meta-material engineering such as negative refractive index.\cite{Valentine08} However, we would like to emphasize that almost same performance results have been demonstrated from other engineered structures consisting purely of lossless dielectric media.\cite{HJ_Chang_10, Notomi00} Recently, Ishizaki and Noda have emphasized the similarity between the SPP wave and the surface state of a 3-D PhC possessing a 3-D PBG.\cite{Ishizaki09}    

In fact, what makes metal {\em metallic} is {\em the presence of negative permittivity} rather than the metallic absorption originated from the imaginary part of the permittivity. One can easily show that any medium with $\epsilon(\omega) < 0$ cannot support any propagating planewave solution.\cite{Jackson_book} This result reminds us of light propagation behavior within the PBG. This observation suggests to us another viewpoint for metal --- metal as a {\em natural} 3-D PBG material. In principle, we can carve and mold metal into an arbitrary geometrical shape. Interestingly, it has been known that even a tiny section (dimension $<$ 10 nm) of metal does not loose its bulk optical properties,\cite{Stockman04} which is in contrast to the case of an {\em artificial} 3-D PhC; at least several lattice periods are needed to function as a PBG material.\cite{Noda00} 

Another aspect of metal is that conduction electrons are strongly coupled to the electromagnetic fields. In fact, plasmons refer to the quantum mechanical eigenstates of coupled electron and photon states.\cite{Kittel_book} In the presence of damping ($\gamma_m > 0$), the energy of plasmons will be converted into heat or mechanical energy. However, {\em the damping mechanism is not an essential aspect of plasmons}. Therefore, we can still think of the total energy of plasmons or {\em the kinetic energy of conduction electrons}. 

\subsection{Definition of plasmonicity}
Here, we would like to find an expression for the kinetic energy density of electrons, $u_{\rm kin} ({\bf r}, t)$, in metal. 
We denote the mass of an electron and the volume density of electrons as $m_e$ and $N_e$, respectively. Then, we can write the time-averaged kinetic energy of electrons per unit volume as 

\begin{equation}
u_{\rm kin}({\bf r},t) \equiv \frac{1}{2} m_e N_e ({\bf r},t) \left< {\bf v}({\bf r},t) \cdot {\bf v}({\bf r},t) \right>_T\label{eq:eq30}
\end{equation}
The above expression can be rewritten in the following way. 
\begin{eqnarray}
u_{\rm kin}({\bf r},t) &=& \frac{1}{2} \frac{m_e}{N_e e^2} \left< (N_e e {\bf v}({\bf r},t)) \cdot (N_e e {\bf v}({\bf r},t)) \right>_T\\
&=& \frac{1}{2} \left[ \frac{m_e}{N_e e^2} \right] \left< {\bf J}({\bf r},t) \cdot {\bf J}({\bf r},t) \right>_T
\end{eqnarray}
In general, the current density vector ${\bf J}({\bf r},t)$ is not a constant in time due to the presence of non-negligible damping. If we assume that damping is negligible within a few optical cycles, we can write ${\bf J}({\bf r},t) \approx (1/2) \left( {\bf \tilde{J}}({\bf r}) \exp(- i \omega t) + {\bf \tilde{J}}^*({\bf r}) \exp(i \omega t) \right)$. 
\begin{eqnarray}
u_{\rm kin}({\bf r}) &\approx& \frac{1}{4} \left[ \frac{m_e}{N_e e^2} \right] |{\bf \tilde{J}}({\bf r})|^2\\
&=& \frac{1}{4} \left[ \frac{m_e}{N_e e^2} \right] |\sigma(\omega)|^2 |{\bf \tilde{E}}({\bf r})|^2
\end{eqnarray}
Here, we have used Ohm's law ${\bf \tilde{J}}({\bf r}) = \sigma(\omega) {\bf \tilde{E}}({\bf r})$. With the Drude model, we have 
\begin{equation}
\sigma(\omega) = \frac{\epsilon_0 \omega_p^2}{\gamma_m  - i \omega},\label{eq:eq21}
\end{equation}
where $\omega_p^2 = N_e e^2/(\epsilon_0 m_e)$. Therefore, we obtain
\begin{equation}
u_{\rm kin}({\bf r}) = \frac{\epsilon_0}{4}  \frac{\omega_p^2}{\gamma_m^2 + \omega^2} |{\bf \tilde{E}}({\bf r})|^2
\end{equation}
We can generalize the above result for the presence of damping such that 
\begin{equation}
u_{\rm kin}({\bf r}, t) = \frac{\epsilon_0}{2}  \frac{\omega_p^2}{\gamma_m^2 + \omega^2} \left< {\bf E}({\bf r},t) \cdot {\bf E}({\bf r},t) \right>_T\label{eq:eq31}
\end{equation}
It is interesting to see that this kinetic energy density term can be found from $u_E$ and $u_M$ (See Eqs.~(\ref{eq:eq6}) and (\ref{eq:eq7})). It should be noted that the role of kinetic energy density or kinetic inductance in developing a circuit model for plasmonics has been emphasized by Staffaroni, {\em et al.}\cite{Yabo_circuit} 

Now, we are in a good position to apply our understanding of $u_{\rm kin}$ to define a new quantitative measure of the degree of plasmonic character. {\em Plasmonicity} $\Pi$ can be defined the fraction of total kinetic energy over the total electromagnetic energy in the limit of zero damping as 
\begin{equation}
\Pi \equiv 2 \times \frac{\rm Total~kinetic~energy}{\rm Total~EM~energy} = 2 \times \frac{\int_{\rm metal} d^3 {\bf r}~u_{\rm kin} ({\bf r})}{\int_{V} d^3 {\bf r}~ \left( u_E ({\bf r}) + u_M ({\bf r}) \right)}~~~~~({\rm in~the~limit}~\gamma_m \rightarrow 0)\label{eq:eq20}
\end{equation}
The reason we take the zero damping limit is that many actual experiments are performed at low temperature.\cite{Min09,Oulton09,Hill_OPEX_09} Even at room temperature, $\gamma_m$ is smaller than $\omega$ by at least one order of magnitude. Moreover, as we have emphasized in the above, {\em the damping or the absorption is not an essential characteristic of metal.} In this way, we can make the definition temperature-independent. The prefactor 2 is to scale up the maximum achievable $\Pi$ to 1, which will be shown in the following examples. 

In the following subsections, we shall examine 1) the SPP mode at a simple metal/dielectric interface and 2) the hybrid metal/PhC nanocavity, with which we shall show the validity of the definition of plasmonicity.

%%%%%%%%%%%%%%%%%%%%%%%%%%%%%%%%%%%%%%%%%%%%%%%%%%%%%%
\begin{figure}%[htb]
\centering\includegraphics[width=16cm]{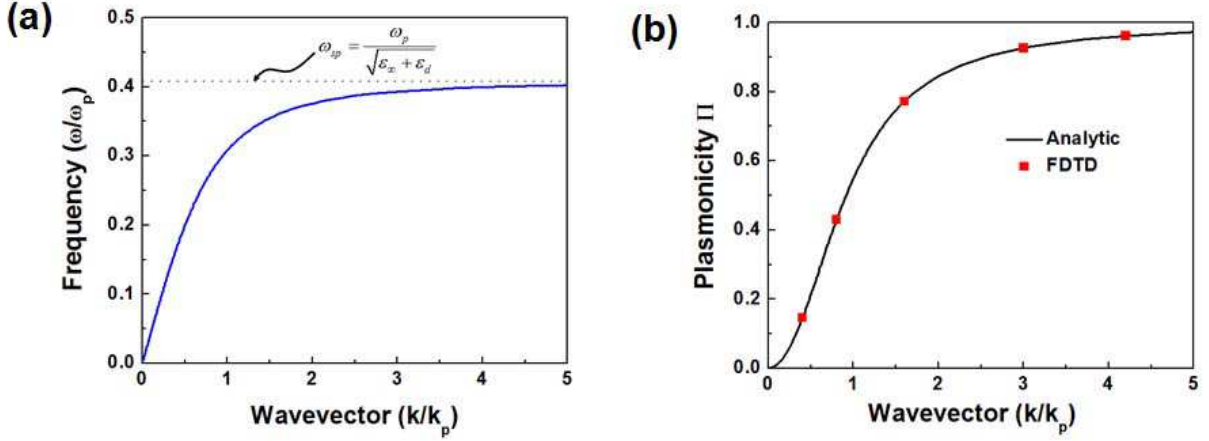}
\caption{\label{fig:fig8}We analyze the simplest surface plasmon polariton mode formed at a dielectric/metal interface. (a) Theoretical dispersion relation ($\omega$-$k$) when $\epsilon_d$ = 5.0 and $\epsilon_{\infty}$ = 1.0, where we assume the Drude model for the metal, $\epsilon_m (\omega) = \epsilon_{\infty} - \omega_p^2 / ( \omega^2 + i \gamma_m \omega )$. (b) Plasmonicity is obtained by analytic calculation (solid curve) and compared with the result obtained using FDTD (square dots).}
\end{figure}
%%%%%%%%%%%%%%%%%%%%%%%%%%%%%%%%%%%%%%%%%%%%%%%%%%%%%%%

\subsection{The simple SPP mode} 
We consider the SPP mode formed at a flat metal/dielectric interface. We assume two semi-infinite media that meet at the $z$ =0 plane; a dielectric medium with $\epsilon_d$ for $z >0$ and the Drude metal with $\epsilon_m(\omega)$ for $z <0$. We also assume that the surface confined SPP mode propagates in the $x$ direction. Then, the dispersion relation of the SPP mode is given by\cite{Pitarke07} 
\begin{equation}
k_x^2 = \left( \frac{\omega}{c} \right) \frac{\epsilon_m \epsilon_d}{ \epsilon_m + \epsilon_d}
\end{equation}

Figure.~\ref{fig:fig8}(a) shows the dispersion curve for $\epsilon_d = 5.0$ and $\epsilon_{\infty} = 1.0$. The wavevector and the frequency are normalized by $k_p = \omega_p /c$ and $\omega_p$, respectively. At small wavevectors, the curve increases almost linearly; in fact, it asymptotically approaches the light line defined by $\omega = c k / \epsilon_d^{1/2}$. At large wavevectors, however, the curve rolls off and saturates at $\omega_{\rm sp} \equiv \omega_p / ( \epsilon_{\infty} + \epsilon_d )^{1/2}$. Traditionally, the SPP mode at this large wavevector limit becomes the classical surface charge density wave and is considered as {\em surface plasmon-like}. In the opposite limit, of course, the mode is {\em photon-like}. Now let us quantify the degree of this plasmonic character. 

Using the definition of $\Pi$ (Eq.~(\ref{eq:eq20})), we have 
\begin{equation}
\Pi_{\rm SPP} = \frac{ \displaystyle \int_0^{1/k_x} dx \int_0^{-\infty} dz ~\epsilon_0 \left( \frac{\omega_p}{\omega}  \right)^2 \left< {\bf E}_m \cdot {\bf E}_m \right>_T }{\displaystyle \int_0^{1/k_x} dx \int_0^{\infty} dz ~\epsilon_0 \epsilon_d  \left< {\bf E}_d \cdot {\bf E}_d \right>_T + \displaystyle \int_0^{1/k_x} dx \int_0^{-\infty} dz ~\epsilon_0 \epsilon_{\infty}  \left< {\bf E}_m \cdot {\bf E}_m \right>_T},
\end{equation}
where ${\bf E}_m$ and ${\bf E}_d$ are electric-fields in metal and dielectrics, respectively. It is not so difficult to carry out the above integration analytically. Thus, we get 
\begin{equation}
\Pi_{\rm SPP} = \frac{ \displaystyle \left( \frac{\omega_p}{\omega} \right)^2 \frac{1}{\epsilon_m^2} }{ \displaystyle \frac{1}{\epsilon_d} + \frac{\epsilon_{\infty}}{\epsilon_m^2} }\label{eq:eq33}
\end{equation}

One can show that $\Pi_{\rm SPP}$ reaches its maximum 1 in the limit of large wavevector $k_x$ where $\omega \rightarrow \omega_p / (\epsilon_{\infty} + \epsilon_d)^{1/2}$ and $\epsilon_m \rightarrow -\epsilon_d$. The solid curve in Fig.~\ref{fig:fig8}(b) shows the analytical result from Eq.~(\ref{eq:eq33}). We have also performed FDTD simulations and the results are overlaid on the same graph, represented by square dots.  

\subsection{Hybrid metal/PhC nanocavity}

%%%%%%%%%%%%%%%%%%%%%%%%%%%%%%%%%%%%%%%%%%%%%%%%%%%%%%%
\begin{figure}%[htb]
\centering\includegraphics[width=16cm]{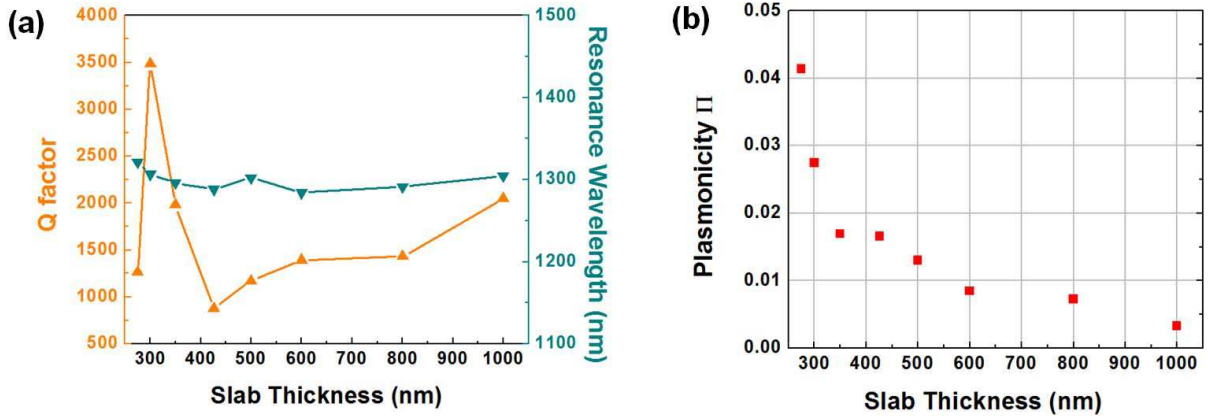}
\caption{\label{fig:fig9}Here, we assume an ideal gold substrate by setting its damping constant to zero. (a) Quality factor of the same dipole mode as shown in Fig.~\ref{fig:fig6}. Since the absorption in gold has been quenched, quality factors relate to radiation losses. Again, we tune the lattice constant ($a$) to approximately keep the emission wavelength at $\sim 1.3 \mu{\rm m}$. (b) The degree of the plasmonic character, `{\em plasmonicity}' of the dipole mode as a function of the slab thickness. }
\end{figure}
%%%%%%%%%%%%%%%%%%%%%%%%%%%%%%%%%%%%%%%%%%%%%%%%%%%%%%%

Now, we return to the hybrid metal/PhC nanocavity and apply the definition of $\Pi$. It should be noted that we set $\gamma_m = 0$ for all calculations in this subsection. First, we calculate $Q$ of the same dipole mode as we vary the slab thickness. The lattice constant has been tuned to keep approximately the same emission wavelength of $\sim 1.3 \mu{\rm m}$. Figure~\ref{fig:fig9}(a) shows $Q$ and the wavelength as a function of the slab thickness. Here, we would like to emphasize that $Q$ obtained in this way is different from the radiation $Q$ defined in Eq.~(\ref{eq:eq13}), because the optical constants of gold have been changed. Though the change is small, it is large enough to produce a noticeable difference in $Q_{\rm rad}$ because the reflectivity of the metal is not the same any more. The presence of a peak in $Q$ at $T = 300$ nm seems to have the similar origin, the vertical resonance within the slab, as argued by Tandaechanurat, {\em et al.}\cite{Tandaechanurat08} Except for this feature, $Q$ tends to increase almost linearly up to 2,000 at $T = 1,000$ nm. 

We plot FDTD simulated $\Pi_{\rm cav}$ values in Fig.~\ref{fig:fig9}(b). When $T > 600$ nm, $\Pi_{\rm cav}$ is below 1\%. Even when $T = 275$ nm, $\Pi_{\rm cav}$ is just above 4\%. These values are much lower compared to those obtained by the SPP mode; only at very small wavevectors of $k/k_p \leq 0.2$, such low values would be obtained. Therefore, our hybrid metal/PhC cavity mode is {\em photonic-like} rather than {\em plasmon-like} and the role of the metal mirror is not to generate plasmons but to serve as a mirror. 

\section{Discussion and Conclusion}

\begin{figure}%[htb]
\centering\includegraphics[width=12cm]{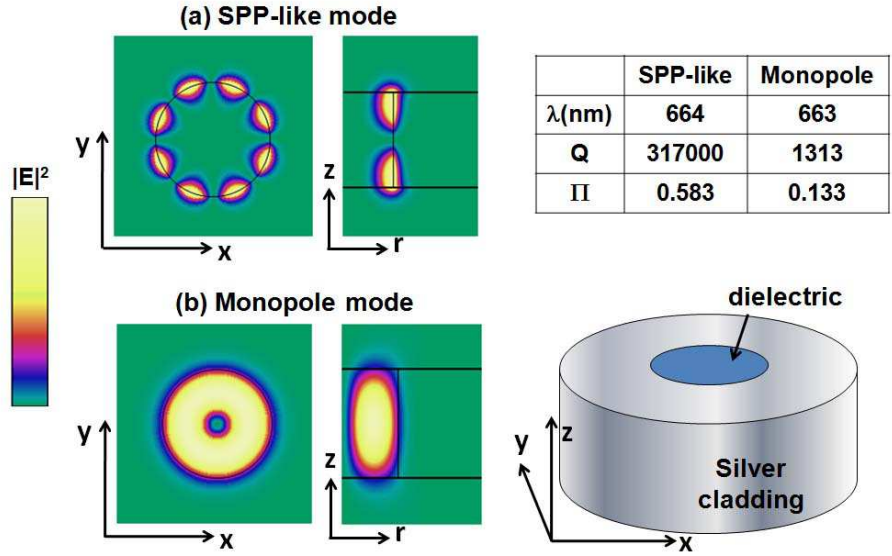}
\caption{\label{fig:fig10}We take an example from our previous paper, a metal-clad nanodisk.\cite{J_Huang_10} The diameter of the dielectric disk is 220 nm. Here, two representative modes are analyzed, (a) one is a SPP-like surface confined mode and (b) the other is a photonic-like monopole mode. Their emission wavelengths, quality factors, and plasmonicities are obtained through FDTD simulations. All values correspond to ideal silver with zero damping.}
\end{figure}

We apply our definition of $\Pi$ to the metal nanodisk cavity, which we have proposed recently.\cite{J_Huang_10} Specifically, we take the two example modes, the monopole mode and the SPP-like mode, both emitting at $\sim 664$ nm (See Fig.~\ref{fig:fig10}). The SPP-like mode, as expected, shows a large $\Pi$ of 58.3\% while the monopole mode shows $\Pi$ of 13.3 \%. Therefore, we believe that the proposed formula provides reasonable quantitative measures for a wide range of metallic nanocavity designs. If metallic and dielectric structures of complex geometrical shapes play an important role in the formation of an electromagnetic mode of interest, a simple judgment based on the the mode profile can be misleading and does not provide any useful information on the internal properties of plasmons. However, the {\em plasmonicty} $\Pi$ requires simple volume integrations of energy densities. Thus, it can be easily applied to various situations regardless of how complex the cavity geometry is.    

There are two different routes of using metallic nanocavities for building coherent light emitters. One is to build an efficient light emitter, where the role of the metal is to serve as a mirror. One such example is our hybrid metal/PhC nanocavity. The other way is to build an efficient plasmonic source, also known as a SPASER (Surface Plasmon Amplification by Stimulated Emission of Radiation).\cite{Stockman03} Such a (dark) plasmonic source would generate more plasmons rather than photons. In this context, it seems that the true SPASER should have very small radiation efficiency. However, as discussed in the previous sections, the conventional definition of radiation efficiency (Eq.~(\ref{eq:eq10})) could be misleading since it contains the effect of the irreversible damping process and the metallic absorption is not an essential property of plasmons. The concept of plasmonicity defined here may provide a useful guideline for the optimization of the SPASER source. By the same token, it could serve equally well for the development of efficient photon sources based on a metallic mirror. In both applications, however, increasing $Q_{\rm tot}$ will be critical to achieve the condition for the stimulated emission.

%%%%%%%%%%%%%%%%%%%%%%%%%%%%%%%%%%%%%%%%%%%%%%%%%%%%%%%
%%%%%%%%%%%%%%%%%%%%%%%%%%%%%%%%%%%%%%%%%%%%%%%%%%%%%%%

\section*{Acknowledgments} 
The authors would like to acknowledge support from the Defense Advanced Research Projects Agency under the Nanoscale Architecture for Coherent Hyperoptical Sources program under grant \#W911NF-07-1-0277 and from the National Science Foundation through NSF CIAN ERC under grant \#EEC-0812072.

%%%%%%%%%%%%%%%%%%%%%%% References %%%%%%%%%%%%%%%%%%%%%%%%%

%\bibliographystyle{osajnl}
%\bibliography{Metal_PhC_hybrid}

\begin{thebibliography}{10}
\newcommand{\enquote}[1]{``#1''}

\bibitem{Purcell46}
E.~M. Purcell, \enquote{Spontaneous emission probabilities at radio
  frequencies,} Phys.\ Rev. \textbf{69}, 681 (1946).

\bibitem{Gerard_LT99}
J.-M. G\'{e}rard and B.~Gayral, \enquote{Strong purcell effect for inas quantum
  boxes in three-dimensional solid-state microcavities,} J. Lightwave Technol.
  \textbf{17}, 2089 --2095 (1999).

\bibitem{Yokoyama92}
H.~Yokoyama, \enquote{Physics and device applications of optical
  microcavities,} Science \textbf{256}, 66--70 (1992).

\bibitem{Yablonovitch87}
E.~Yablonovitch, \enquote{Inhibited spontaneous emission in solid-state physics
  and electronics,} Phys. Rev. Lett. \textbf{58}, 2059--2062 (1987).

\bibitem{Sajeev87}
S.~John, \enquote{Strong localization of photons in certain disordered
  dielectric superlattices,} Phys. Rev. Lett. \textbf{58}, 2486--2489 (1987).

\bibitem{Joannopoulos_book}
J.~D. Joannopoulos, S.~G. Johnson, J.~N. Winn, and R.~D. Meade, \emph{Photonic
  Crystals: Molding the Flow of Light} (Princeton University Press, Princeton,
  NJ, 2008), 2nd ed.

\bibitem{Yeh76}
P.~Yeh, A.~Yariv, and C.~S. Hong, \enquote{Electromagnetic propagation in
  periodic stratified media. {I.} {G}eneral theory,} \josa \textbf{67},
  423--438 (1976).

\bibitem{Jewell91}
J.~L. Jewell, J.~P. Harbison, A.~Scherer, Y.~H. Lee, and L.~T. Florez,
  \enquote{Vertical-cavity surface-emitting lasers: Design, growth,
  fabrication, characterization,} \jqe \textbf{27}, 1332--1346 (1991).

\bibitem{O_Painter_99}
O.~Painter, R.~K. Lee, A.~Yariv, A.~Scherer, J.~D. O'Brien, P.~D. Dapkus, and
  I.~Kim, \enquote{Two-dimensional photonic band-gap defect mode laser,}
  Science \textbf{284}, {1819--1821} (1999).

\bibitem{H_Y_Ryu_03}
H.-Y. Ryu, M.~Notomi, and Y.-H. Lee, \enquote{High-quality-factor and
  small-mode-volume hexapole modes in photonic-crystal-slab nanocavities,}
  Appl. Phys. Lett. \textbf{83}, 4294--4296 (2003).

\bibitem{BSSong05}
B.~S. Song, S.~Noda, T.~Asano, and Y.~Akahane, \enquote{Utra-high-{Q} photonic
  double-heterostructure nanocavity,} Nat. Mater. \textbf{4}, 207--210 (2005).

\bibitem{Deotare09}
P.~B. Deotare, M.~W. McCutcheon, I.~W. Frank, M.~Khan, and M.~Loncar,
  \enquote{High quality factor photonic crystal nanobeam cavities,} Appl. Phys.
  Lett. \textbf{94}, 121106 (2009).

\bibitem{Painter99}
O.~Painter, J.~Vu\v{c}kovi\v{c}, and A.~Scherer, \enquote{Defect modes of a
  two-dimensional photonic crystal in an optically thin dielectric slab,} J.
  Opt. Soc. Am. B \textbf{16}, 275--285 (1999).

\bibitem{S_G_Johnson_99}
S.~G. Johnson, S.~Fan, P.~R. Villeneuve, J.~D. Joannopoulos, and L.~A.
  Kolodziejski, \enquote{Guided modes in photonic crystal slabs,} Phys. Rev. B
  \textbf{60}, 5751--5758 (1999).

\bibitem{Kim_PRB_06}
S.-H. Kim, S.-K. Kim, and Y.-H. Lee, \enquote{Vertical beaming of
  wavelength-scale photonic crystal resonators,} Physical Review B \textbf{73},
  235117 (2006).

\bibitem{Khankhoje10}
U.~K. Khankhoje, S.-H. Kim, B.~C. Richards, J.~Hendrickson, J.~Sweet, J.~D.
  Olitzky, G.~Khitrova, H.~M. Gibbs, and A.~Scherer, \enquote{Modelling and
  fabrication of gaas photonic-crystal cavities for cavity quantum
  electrodynamics,} Nanotechnology \textbf{21}, 065202 (2010).

\bibitem{Hinds}
E.~A. Hinds, \emph{in Cavity Quantum Electrodynamics} (Academic Press, Inc,
  Orlando, 1994). Edited by P. R. Berman.

\bibitem{S_H_Kim_07a}
S.-H. Kim, M.-K. Seo, J.-Y. Kim, and Y.-H. Lee, \enquote{Effects of a bottom
  substrate on emission properties of a photonic crystal nanolaser,} Indium
  Phosphide \& Related Materials, 2007. IPRM '07. IEEE 19th International
  Conference on pp. 480--483 (2007).

\bibitem{Toishi09}
M.~Toishi, D.~Englund, A.~Faraon, and J.~Vu\v{c}kovi\'{c},
  \enquote{High-brightness single photon source from a quantum dot in a
  directional-emission nanocavity,} Opt. Express \textbf{17}, 14618--14626
  (2009).

\bibitem{H_G_Park_04}
H.-G. Park, S.-H. Kim, S.-H. Kwon, Y.-G. Ju, J.-K. Yang, J.-H. Baek, S.-B. Kim,
  and Y.-H. Lee, \enquote{Electrically driven single-cell photonic crystal
  laser,} Science \textbf{305}, {1444--1447} (2004).

\bibitem{MKSeo07}
M.-K. Seo, K.-Y. Jeong, J.-K. Yang, Y.-H. Lee, H.-G. Park, and S.-B. Kim,
  \enquote{Low threshold current single-cell hexapole mode photonic crystal
  laser,} Appl. Phys. Lett. \textbf{90}, 171122 (2007).

\bibitem{Okumura09}
T.~Okumura, M.~Kurokawa, M.~Shirao, D.~Kondo, H.~Ito, N.~Nishiyama,
  T.~Maruyama, and S.~Arai, \enquote{Lateral current injection {GaInAsP}/{InP}
  laser on semi-insulating substrate for membrane-based photonic circuits,}
  \opex \textbf{17}, 12564--12570 (2009).

\bibitem{Ellis11}
B.~Ellis, M.~A. Mayer, G.~Shambat, T.~Sarmiento, J.~Harris, E.~E. Haller, and
  J.~Vuckovic, \enquote{Ultralow-threshold electrically pumped quantum-dot
  photonic-crystal nanocavity laser,} Nat. Photon. \textbf{5}, 297--300 (2011).

\bibitem{Choquette10}
C.~M. Long, A.~V. Giannopoulos, and K.~D. Choquette, \enquote{Lateral current
  injection photonic crystal membrane light emitting diodes,} J. Vac. Sci.
  Technol. B \textbf{28}, 359--364 (2010).

\bibitem{Nozaki07}
K.~Nozaki, S.~Kita, and T.~Baba, \enquote{Room temperature continuous wave
  operation and controlled spontaneous emission in ultrasmall photonic crystal
  nanolaser,} Opt. Express \textbf{15}, 7506--7514 (2007).

\bibitem{Kim_ICTON_09}
S.-H. Kim, Y.-H. Lee, J.~Huang, and A.~Scherer, \enquote{Unidirectional
  vertical emission from photonic crystal nanolasers,} in \enquote{Transparent
  Optical Networks, 2009. ICTON '09. 11th International Conference on,}
  (2009), pp. 1--4.

\bibitem{M_L_Povinelli_03}
M.~L. Povinelli, S.~G. Johnson, J.~D. Joannopoulos, and J.~B. Pendry,
  \enquote{Toward photonic-crystal metamaterials: Creating magnetic emitters in
  photonic crystals,} Appl.\ Phys.\ Lett. \textbf{82}, {1069--1071} (2003).

\bibitem{Johnson72}
P.~B. Johnson and R.~W. Christy, \enquote{Optical constants of the noble
  metals,} Phys. Rev. B \textbf{6}, 4370--4379 (1972).

\bibitem{Fan02}
S.~Fan and J.~D. Joannopoulos, \enquote{Analysis of guided resonances in
  photonic crystal slabs,} Phys. Rev. B \textbf{65}, 235112 (2002).

\bibitem{Dowling91}
J.~P. Dowling, M.~O. Scully, and F.~DeMartini, \enquote{Radiation pattern of a
  classical dipole in a cavity,} Opt. Commun. \textbf{82}, 415--419 (1991).

\bibitem{Bakir06}
B.~B. Bakir, C.~Seassal, X.~Letartre, P.~Viktorovitch, M.~Zussy, L.~D. Cioccio,
  and J.~M. Fedeli, \enquote{Surface-emitting microlaser combining
  two-dimensional photonic crystal membrane and vertical bragg mirror,} Appl.
  Phys. Lett. \textbf{88}, 081113 (2006).

\bibitem{Ochiai01}
T.~Ochiai and K.~Sakoda, \enquote{Dispersion relation and optical transmittance
  of a hexagonal photonic crystal slab,} Phys. Rev. B \textbf{63}, 125107
  (2001).

\bibitem{J_Kang_09}
J.-H. Kang, M.-K. Seo, S.-K. Kim, S.-H. Kim, M.-K. Kim, H.-G. Park, K.-S. Kim,
  and Y.-H. Lee, \enquote{Polarized vertical beaming of an engineered hexapole
  mode laser,} Opt. Express \textbf{17}, 6074--6081 (2009).

\bibitem{Min09}
B.~Min, E.~Ostby, V.~Sorger, E.~Ulin-Avila, L.~Yang, X.~Zhang, and K.~Vahala,
  \enquote{High-{Q} surface-plasmon-polarition whispering-gallery microcavity,}
  \nat \textbf{457}, 455--459 (2009).

\bibitem{Oulton09}
R.~F. Oulton, V.~J. Sorger, T.~Zentgaraf, R.~M. Ma, C.~Gladden, L.~Dai,
  G.~Bartal, and X.~Zhang, \enquote{Plasmon lasers at deep subwavelength
  scale,} \nat \textbf{457}, 629--632 (2009).

\bibitem{Seo_NL09}
M.-K. Seo, S.-H. Kwon, H.-S. Ee, and H.-G. Park, \enquote{Full
  three-dimensional subwavelength high-{Q} surface-plasmon-polariton cavity,}
  Nano Lett. \textbf{9}, 4078--4082 (2009).

\bibitem{Hill_OPEX_09}
M.~T. Hill, M.~Marell, E.~S.~P. Leong, B.~Smalbrugge, Y.~Zhu, M.~Sun, P.~J. van
  Veldhoven, E.~J. Geluk, F.~Karouta, Y.-S. Oei, R.~N\"{o}tzel, C.-Z. Ning, and
  M.~K. Smit, \enquote{Lasing in metal-insulator-metal sub-wavelength plasmonic
  waveguides,} \opex \textbf{17}, 11107--11112 (2009).

\bibitem{J_Huang_10}
J.~Huang, S.-H. Kim, and A.~Scherer, \enquote{Design of a surface-emitting,
  subwavelength metal-clad disk laser in the visible spectrum,} \opex
  \textbf{18}, 19581--19591 (2010).

\bibitem{Hill_JOSAB_10}
M.~T. Hill, \enquote{Status and prospects for metallic and plasmonic
  nano-lasers,} \josab \textbf{27}, B36--B44 (2010).

\bibitem{Chang09_QE}
S.-W. Chang and S.~L. Chuang, \enquote{Fundamental formulation for plasmonic
  nanolasers,} Quantum Electronics, IEEE Journal of \textbf{45}, 1014 --1023
  (2009).

\bibitem{Chang09_OL}
S.-W. Chang and S.~L. Chuang, \enquote{Normal modes for plasmonic nanolasers
  with dispersive and inhomogeneous media,} Opt. Lett. \textbf{34}, 91--93
  (2009).

\bibitem{Jackson_book}
J.~D. Jackson, \emph{Classical Electrodynamics} (Wiley, New York, 1998), 3rd
  ed.

\bibitem{Tandaechanurat08}
A.~Tandaechanurat, S.~Iwamoto, M.~Nomura, N.~Kumagai, and Y.~Arakawa,
  \enquote{Increase of {Q}-factor in photonic crystal {H1}-defect nanocavities
  after closing of photonic bandgap with optimal slab thickness,} \opex
  \textbf{16}, 448--455 (2008).

\bibitem{Mizrahi_08}
A.~Mizrahi, V.~Lomakin, B.~A. Slutsky, M.~P. Nezhad, L.~Feng, and Y.~Fainman,
  \enquote{Low threshold gain metal coated laser nanoresonators,} Opt. Lett.
  \textbf{33}, 1261--1263 (2008).

\bibitem{Ibanescu05}
M.~Ibanescu, S.~G. Johnson, D.~Roundy, Y.~Fink, and J.~D. Joannopoulos,
  \enquote{Microcavity confinement based on an anomalous zero group-velocity
  waveguide mode,} \ol \textbf{30}, 552--554 (2005).

\bibitem{SHKim_unplished}
S.-H. Kim, J.~Huang, and A.~Scherer {a}re preparing a manuscript to be called
  {``}A photonic-crystal nanocavity laser in a very thick-slab{"}.

\bibitem{Taillaert02}
D.~Taillaert, W.~Bogaerts, P.~Bienstman, T.~Krauss, P.~Van~Daele, I.~Moerman,
  S.~Verstuyft, K.~De~Mesel, and R.~Baets, \enquote{An out-of-plane grating
  coupler for efficient butt-coupling between compact planar waveguides and
  single-mode fibers,} IEEE Quantum Electron. \textbf{38}, 949 --955 (2002).

\bibitem{Fujita99}
M.~Fujita, A.~Sakai, and T.~Baba, \enquote{Ultrasmall and ultralow threshold
  {GaInAsP-InP} microdisk injection lasers: design, fabrication, lasing
  characteristics, and spontaneous emission factor,} IEEE Sel. Top. Quantum
  Electron. \textbf{5}, 673 --681 (1999).

\bibitem{Friedler09}
I.~Friedler, C.~Sauvan, J.~P. Hugonin, P.~Lalanne, J.~Claudon, and J.~M.
  G\'{e}rard, \enquote{Solid-state single photon sources: the nanowire
  antenna,} \opex \textbf{17}, 2095--2110 (2009).

\bibitem{Li11}
W.-D. Li, F.~Ding, J.~Hu, and S.~Y. Chou, \enquote{Three-dimensional cavity
  nanoantenna coupled plasmonic nanodots for ultrahigh and uniform
  surface-enhanced raman scattering over large area,} Opt. Express \textbf{19},
  3925--3936 (2011).

\bibitem{Valentine08}
J.~Valentine, S.~Zhang, T.~Zentgraf, E.~Ulin-Avila, D.~A. Genov, G.~Bartal, and
  X.~Zhang, \enquote{Three-dimensional optical metamaterial with a negative
  refractive index,} \nat \textbf{455}, 376--379 (2008).

\bibitem{HJ_Chang_10}
H.-J. Chang, S.-H. Kim, Y.-H. Lee, E.~P. Kartalov, and A.~Scherer, \enquote{A
  photonic-crystal optical antenna for extremely large local-field
  enhancement,} Opt. Express \textbf{18}, 24163--24177 (2010).

\bibitem{Notomi00}
M.~Notomi, \enquote{Theory of light propagation in strongly modulated photonic
  crystals: Refractionlike behavior in the vicinity of the photonic band gap,}
  Phys. Rev. B \textbf{62}, 10696--10705 (2000).

\bibitem{Ishizaki09}
K.~Ishizaki and S.~Noda, \enquote{Manipulation of photons at the surface of
  three-dimensional photonic crystals,} \nat \textbf{460}, 367--371 (2009).

\bibitem{Stockman04}
M.~I. Stockman, \enquote{Nanofocusing of optical energy in tapered plasmonic
  waveguides,} Phys. Rev. Lett. \textbf{93}, 137404 (2004).

\bibitem{Noda00}
S.~Noda, K.~Tomoda, N.~Yamamoto, and A.~Chutinan, \enquote{Full
  three-dimensional photonic bandgap crystals at near-infrared wavelengths,}
  Science \textbf{289}, 604--606 (2000).

\bibitem{Kittel_book}
C.~Kittel, \emph{Introduction to solid state physics} (John Wiley \& Sons,
  Inc., 2005), eighth ed.

\bibitem{Yabo_circuit}
M.~Staffaroni, J.~Conway, S.~Vedantam, J.~Tang, and E.~Yablonovitch,
  \enquote{Circuit analysis in metal-optics,}
  {h}ttp://arxiv.org/abs/1006.3126.

\bibitem{Pitarke07}
J.~M. Pitarke, V.~M. Silkin, E.~V. Chulkov, and P.~M. Echenique,
  \enquote{Theory of surface plasmons and surface-plasmon polaritions,} Rep.
  Prog. Phys. \textbf{70}, 1--87 (2007).

\bibitem{Stockman03}
D.~J. Bergman and M.~I. Stockman, \enquote{Surface plasmon amplification by
  stimulated emission of radiation: Quantum generation of coherent surface
  plasmons in nanosystems,} Phys. Rev. Lett. \textbf{90}, 027402 (2003).

\end{thebibliography}

\newpage

\end{document}